\documentclass{aastex631}

\newcommand{\MM}{M_{\bullet} - M_{\star}}
\newcommand{\Mbh}{M_{\bullet}}
\newcommand{\Mstar}{M_{\star}}

\accepted{May 31, 2024}

\submitjournal{Research Notes of the AAS}

\begin{document}

\title{No Significant Redshift Evolution in the Intrinsic Scatter \\ of the $\MM$ Relation for Overmassive Black Holes}

\correspondingauthor{Fabio Pacucci}
\email{fabio.pacucci@cfa.harvard.edu}

\author[0009-0009-7258-1637]{Carl Audric Guia}
\affiliation{Harvard College, Cambridge, MA 02138, USA}

\author[0000-0001-9879-7780]{Fabio Pacucci}
\affiliation{Center for Astrophysics $\vert$ Harvard \& Smithsonian, Cambridge, MA 02138, USA}
\affiliation{Black Hole Initiative, Harvard University, Cambridge, MA 02138, USA}

\begin{abstract}
In the local Universe, the ratio between the mass of a central black hole and the stellar mass of its host galaxy is $\sim 0.1 \%$. Recently, JWST discovered numerous galaxies at $z>4$ that seem to deviate from the local relation, with black holes overmassive by $10-100$ times. Similar galaxies were also discovered at cosmic noon. The intrinsic scatter in the relation describes how much the evolutionary histories of the single galaxies deviate from the mean evolutionary pattern of their parent dataset. This Research Note examines whether a cosmic evolution of the intrinsic scatter can be detected by assessing its value for datasets in various redshift ranges. Using data from the local Universe ($z < 0.055$), low ($ 0.4 \leq z \leq 0.9$), intermediate ($ 0.9 \leq z \leq 4$), and high ($z > 4$) redshift, we conclude that there is no statistically significant redshift evolution of the intrinsic scatter. 
\end{abstract}

\vspace{0.3cm}
\section{Introduction} \label{sec: intro}
Profound relations link the mass of the central black holes to some physical properties of their host galaxies (see, e.g., \citealt{Magorrian_1998, Ferrarese_Merritt_2000, Gebhardt_2000}), indicating that these two classes of astrophysical objects co-evolved across cosmic time. In the local Universe, these relations suggest that the ratio between the black hole mass $\Mbh$ and the stellar mass $\Mstar$ is  $\sim 0.1 \%$ \citep{reines15}.

Based on new datasets from JWST, \cite{pacucci23} found that galaxies at $z \sim 4 - 7$ deviated from the local relation by $>3 \sigma$, suggesting that black holes are overmassive by a factor in the range $10-100$. If this effect is not due to systematic errors in determining $\Mbh$ and $\Mstar$, then it has profound physical consequences. For instance, it could be suggestive that heavy seeding occurred in the high-$z$ Universe, as several recent studies (see, e.g., \citealt{Scoggins_2023}) showed in detail.

In this Research Note, we investigate the redshift evolution of the intrinsic scatter in the $\MM$ relation. Physically, the scatter indicates how the single evolutionary histories of the galaxies are different from the mean evolutionary pattern. In particular, we assess how the intrinsic scatter computed for higher-redshift datasets of overmassive black holes compares with the reference intrinsic scatter in the local Universe and if a statistically significant evolution can be observed. Because of its physical meaning, assessing whether the intrinsic scatter changes significantly over time may help in understanding if these different datasets of overmassive black holes belong, in reality, to a single population of astrophysical objects.

\section{Data} \label{sec:data}
Data spanning four different redshift regimes are used: the local Universe ($z < 0.055$), low ($ 0.4 \leq z \leq 0.9$), intermediate ($ 0.9 \leq z \leq 4$), and high ($z > 4$) redshift.

For the local Universe sample, we refer to \cite{reines15}, where the authors used 244 broad-line AGNs at $z < 0.055$ and found a best-fit intrinsic scatter of $0.47$ dex. 
This sample is used as a reference for the local Universe to compare with the other datasets, which are instead focused on overmassive black holes at higher redshifts.

Data collected by \cite{mezcua23} is used for the low redshift regime, consisting of 7 systems in the $ 0.4 \leq z \leq 0.9$ range. The intermediate-redshift regime contains 12 new galaxies recently discovered by \cite{mezcua24} at cosmic noon. The high redshift regimes have data coming from a variety of data sets: 8 galaxies from \cite{harikane23}, 12 from \cite{Maiolino_2023}, 1 from \cite{ubler23}, 5 from \cite{stone24}, 1 from \cite{furtak23}, 1 from \cite{kokorev23}, 6 from \cite{yue24}, and 1 from \cite{bogdan24}.

\begin{figure}
    \centering
    \includegraphics[width=0.9\textwidth]{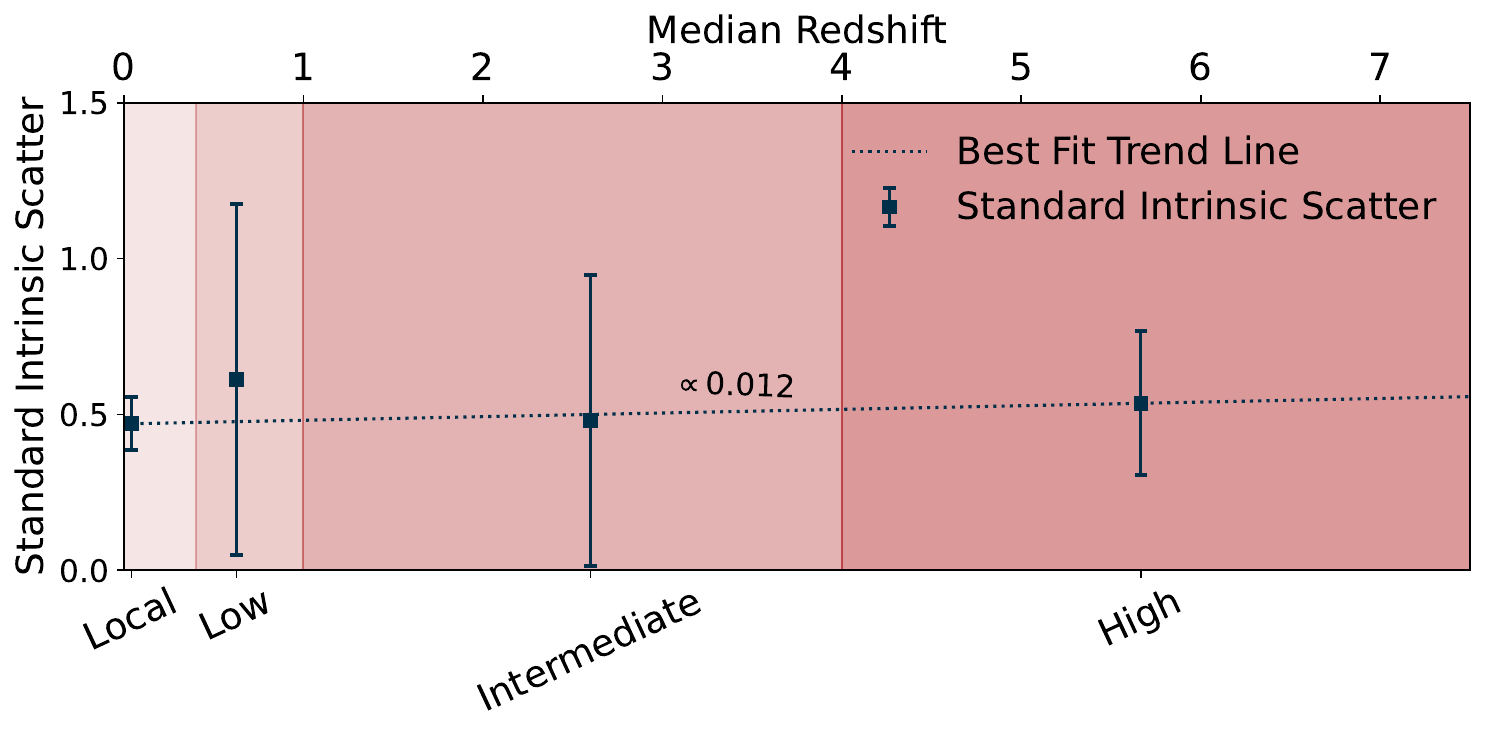}
    \caption{Standard intrinsic scatter as a function of the median redshift for all our datasets in various redshift regimes. Typical values are within the range of $0.45-0.65$ dex. No statistically significant redshift evolution is observed.}
    \label{fig: scatterVSredshift}
\end{figure}

\section{Methods} \label{sec:methods}

We use the same model developed in \cite{pacucci23}, to which the interested reader is referred for details.
An MCMC algorithm, which properly corrects for observational biases, infers the $\MM$ relation of a sample of galaxies and estimates their intrinsic scatter. The parameters that describe the inferred $\MM$ relation are $b$, $m$, and $\nu$. The first two describe the relation between $\Mbh$ and $\Mstar$ in the equation:

\begin{equation}
    \log \left(\frac{\Mbh}{M_{\odot}}\right) = b + m \log \left(\frac{\Mstar}{M_{\odot}}\right) \, .
\end{equation}

In addition, $\sqrt{\nu(1+m^2)}$ is the standard intrinsic scatter in the direction of $\Mbh$. 
The parameters of the MCMC model (e.g., the stellar mass function and the telescope's sensitivity to the detection of a given black hole mass) are adjusted to the median redshift of each dataset. In particular, for the stellar mass function, we used the parameters provided by \cite{Santini_2012} for $z < 4$ and by \cite{song16} for $z > 4$.

Because some data report only upper limits on the $\Mstar$ values, a lower bound was assigned conservatively to the minimum $\Mstar$ of the entire dataset. The asymmetry of the upper and lower bars for some data points indicate non-Gaussian uncertainties; hence, the larger error bar is chosen to model the data as Gaussian, simplifying the statistical analysis. 

\vspace{0.5cm}

\section{Results} \label{sec:results}
To investigate the potential redshift evolution of the standard intrinsic scatter, we characterized each dataset with two numbers: (i) its median redshift and (ii) its inferred standard intrinsic scatter. These latter values are weighted by the number of data points available in each redshift regime in order to infer the best-fit trend line (see Fig. \ref{fig: scatterVSredshift}) using the reduced chi-squared method. The uncertainties associated with the standard intrinsic scatters are derived from the posterior distributions for $\nu$ and $m$, which are directly provided by the MCMC algorithm. Our results are shown in Fig. \ref{fig: scatterVSredshift}.

All values of the standard intrinsic scatter are within the range of $0.45-0.65$ dex.
Visually, there seems to be no significant redshift evolution of the standard intrinsic scatter with redshift, given the considerable uncertainties. We used the Pearson correlation coefficient to test the strength of the observed correlation between redshift and intrinsic scatter to quantify these findings. This analysis yielded a Pearson coefficient of $0.003$ with a p-value of $0.997$. Hence, \textit{current data is insufficient to conclude that there is a significant redshift evolution of the intrinsic scatter}.
Additional data and a better understanding of the uncertainty associated with stellar and black hole mass measurements are thus urgent in addressing whether these populations of overmassive black holes across cosmic times are statistically compatible.

\begin{acknowledgments}
    This project is funded by the Harvard College Research Program for the Spring 2024 semester. 
\end{acknowledgments} 

\software{\texttt{Astropy} \citep{astropy}, \texttt{emcee} \citep{foreman13}.}

\bibliography{ms}{}
\bibliographystyle{aasjournal}

\end{document}